\documentclass[aps,prl,twocolumn,
preprintnumbers,amsmath,amssymb,showpacs,superscriptaddress,amsfonts]{revtex4-1}

 \usepackage{graphicx}
 \usepackage{dcolumn}
 \usepackage{bm}
 \usepackage{color}
 \usepackage{epstopdf}
\renewcommand{\narrowtext}{\begin{multicols}{2} \global\columnwidth20.5pc}
\renewcommand{\v}[1]{{\bf #1}}

\def\gapp{\lower.35em\hbox{$\stackrel{\textstyle>}{\sim}$}}
\def\lapp{\lower.35em\hbox{$\stackrel{\textstyle<}{\sim}$}}
\newcommand{\Eq}[1]{Eq.~(\ref{#1})}
\def\be{\begin{eqnarray}}
\def\ee{\end{eqnarray}}
\begin{document}


\title{The emergence of topologically protected surface states in epitaxial Bi(111) thin films}


\author{Kai Zhu}
\author{Lin Wu}
\author{Xinxin Gong}
\author{Shunhao Xiao}
\author{Shiyan Li}
\author{Xiaofeng Jin}
\email[Corresponding author, E-mail:]{xfjin@fudan.edu.cn}
\affiliation{Department of Physics, State Key Laboratory of Surface Physics, Fudan University, Shanghai 200433,China}
\author{Mengyu Yao}
\author{Dong Qian}
\affiliation{Department of Physics and Astronomy, Key Laboratory of Artificial Structures and Quantum Control, Shanghai Jiao Tong University, Shanghai 200240, China}
\author{Meng Wu}
\author{Ji Feng}
\author{Qian Niu}
\affiliation{International Center for Quantum Materials, School of Physics, Peking University, Beijing 100871, China}
\affiliation{Collaborative Innovation Center of Quantum Matter, Beijing, China}
\author{Fernando de Juan}
\author{Dung-Hai Lee}
\affiliation{Materials Science Division, Lawrence Berkeley National Laboratories, Berkeley, CA
94720}
\affiliation{Department of Physics, University of California at Berkeley, Berkeley, CA94720, USA}


\date{\today}

\begin{abstract}
Quantum transport measurements including the Altshuler-Aronov-Spivak (AAS) and Aharonov-Bohm (AB) effects, universal conductance fluctuations (UCF), and weak anti-localization (WAL) have been carried out on epitaxial Bi thin films ($10-70$ bilayers) on Si(111). The results show that while the film interior is insulating all six surfaces of the Bi thin films are robustly metallic. We propose that these properties are the manifestation of a novel phenomenon, namely, a topologically trivial bulk system can become topologically non-trivial when it is made into a thin film.
We stress that what's observed here is entirely different from the predicted 2D topological insulating state in a single bilayer Bi where only the four side surfaces should possess topologically protected gapless states.
\end{abstract}

\pacs{73.20-r, 73.25.+i, 73.50.-h}

\maketitle

Historically Bi is a material where many interesting condensed matter phenomena were (first) observed. This includes the de Haas-van Alphen effect \cite{SDH}, quantum size confinement effect \cite{QSE}, quantum linear magnetoresistance \cite{QLM}, peculiar superconductivity \cite{SC}, and possibly fractional quantum Hall effect \cite{FQH}. Despite of decades of studies, many fundamental properties of Bi remain poorly understood. For example, the Bi surfaces conduct much better than the bulk \cite{Rev}. Even more dramatically, the surface of Bi remains metallic after the bulk is turned into insulating by reducing the thickness of Bi films
\cite{XSH}.

The last phenomenon is reminiscent to the behavior of topological insulators. Given that many Bi-based compounds (e.g., Bi$_{x}$Sb$_{1-x}$, Bi$_{2}$Se$_{3}$, Bi$_{2}$Te$_{3}$) are topological insulators \cite{TI1,TI2}, ``can reduction of thickness turn the Bi thin film into topological insulator from its topologically trivial bulk ?'' This is the question  we address in this letter. In the following we report the results of a number of quantum transport measurements and theoretical calculations. Combining these investigations we arrive at the the conclusion that Bi thin films are interesting examples where a topologically trivial system becomes non-trivial solely due to the reduction of thickness. Throughout this letter we focus on epitaxial Bi thin films grown on Si(111) substrate.

Polished single crystal semi-insulating Si(111) substrates with high resistivity ($> 10^{6}$ $\Omega$cm at 250~K) were chemically cleaned before put into the ultrahigh vacuum (UHV) chamber. It is  further flashed to 1300~K to obtain a well-ordered ($7\times7$) reconstructed surface, on which the epitaxial growth of Bi is carried out at 300 K \cite{Bi}. The details of our experimental set-up can be found in Ref \cite{Tian,Yin,XSH,AHE}. A representative reflection high energy electron diffraction (RHEED) pattern of our MBE grown thin films is given in the inset of Fig. 1(a). Both large angle and grazing angle x-ray diffraction measurements confirm that the Bi films investigated in this work are fully strain released with bulk lattice constants. Before the film is taken out of the UHV chamber we cap it  with 5~nm MgO to protect it from oxidation. The transport measurements were carried out in an Oxford Cryofree magnet system with temperature down to 1.4~K and magnetic field up to 9~T.

Fig. 1(a) plots the longitudinal resistivity versus temperature. Two characteristic behaviors are observed: a low temperature  metallic behavior ($d\rho_{xx}/dT>0$) and a higher temperature insulating behavior ($d\rho_{xx}/dT<0$). This result is consistent with the behavior of a thin film with an insulating interior but a metallic surface \cite{XSH}. More specifically we attribute the linearly rising resistivity at low temperatures to  the transport behavior of the metallic surface (it is similar to the resistivity behavior of a monolayer indium film \cite{In}). In contrast we attribute the  high temperature behavior to the conduction due to thermally activated (across the interior energy gap) electrons and holes. Additional support of the low temperature conduction being due to the surface states comes from the fact that the low temperature total conductance of Bi films grown on two different substrates stays constant for thickness in the range of 15-27.5 nm. To our knowledge, due to the presence of bulk carriers, none of the as grown topological insulators Bi$_{x}$Sb$_{1-x}$, Bi$_{2}$Se$_{3}$, and Bi$_{2}$Te$_{3}$ has exhibited the insulating to metallic crossover as shown in Fig. 1(a).

\begin{figure}
\includegraphics[width=7cm]{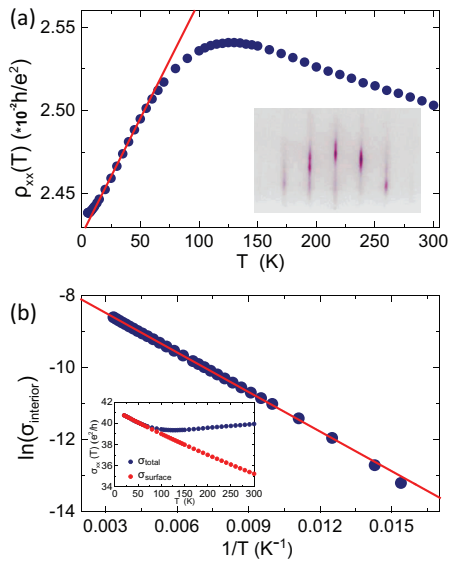}
\caption{\label{fig1}  (a) 4~nm Bi film on Si(111): the resistivity as a function of temperature, and its representative RHEED pattern (inset). (b) The linear relationship between $\ln(\sigma_{interior})$ and $1/T$.}
\end{figure}

To separate the surface and interior conduction we first fit the linearly increasing resistivity by $\rho_{surface}(T)=\rho_{surface}(0)+\kappa$$T$ as marked by the red line in Fig. 1(a). The corresponding surface conductivity is shown in the inset of Fig. 1(b). By subtracting it from the total conductivity we obtain the temperature dependent conductivity of the film interior $\sigma_{interior}(T)$. In Fig.1(b) we plot $\ln(\sigma_{interior}(T))$ versus $1/T$. The slope
gives an estimate of the interior energy gap. The result of this estimate is $\sim$62~meV. Our own angle resolved photoemission spectroscopic (ARPES) data on a 30~nm (75~BL)  Bi(111) film confirms the existence of an indirect energy gap in the film interior\cite{Yao}. Due to the 62 meV gap
the low temperature transport discussed below is completely dominated by the surface conduction.

\begin{figure}
\includegraphics[width=9.0cm]{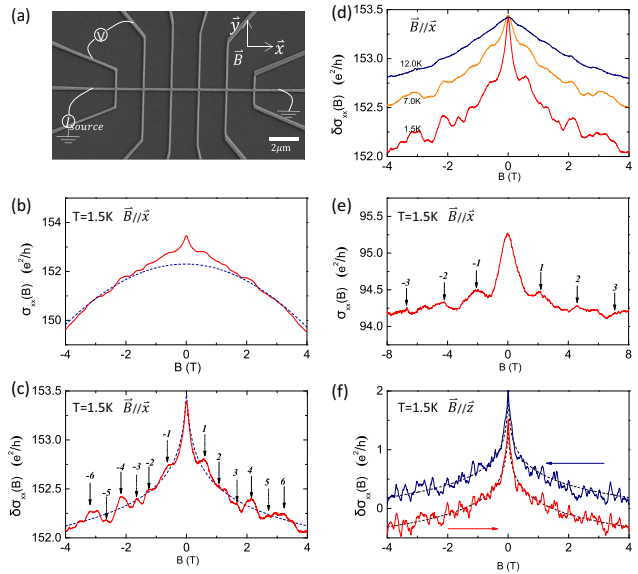}
\caption{\label{fig2} (a) SEM picture of the Bi film device, with a length of 18~$\mu$m, width 220.7~nm, and thickness 16.0~nm. (b) Magneto-conductivity for 16.0~nm thick Bi, measured at 1.5~K with the magnetic field applied along the current direction. (c) Identified AAS and AB oscillations. (d) Temperature dependence of the AAS and AB effects. (e) Dominated AB oscillations for 8.0~nm Bi. (f) UCF with the magnetic field applied perpendicular to Bi film plane.}
\end{figure}

Next we demonstrate all six (top, bottom and four side) surfaces of the film are metallic.   Fig. 2(a) shows a scanning electron microscope (SEM) picture of our device, in which both the current and magnetic field are applied along the x direction. The four probe method is used to measure the electric conductance. Fig. 2(b) shows the raw conductance versus magnetic field data for a 16~nm Bi film at T=1.5~K . Besides the weak anti-localization peak around zero magnetic field reproducible fine structures  are observed on top of a smooth background (marked  by the dashed line). After a background subtraction, quasi-periodic oscillations with a magnetic field period 0.55~T are clearly identified (see Fig. 2(c)). We attribute such oscillation to the  Altshuler-Aronov-Spivak (AAS)  $h/2e$ oscillation \cite{AAS}. Using $h/2e$ as the magnetic flux period, the above mentioned 0.55~T magnetic field period corresponds to a cross-section area of $3.8\times10^{-15}m^{2}$, which is in excellent agreement with the  measured cross-section area $S=w\cdot d=3.5\times10^{-15}m^{2}$ ($w$ and $d$ are the width and the thickness of the film). As usual the amplitudes of the AAS oscillations decreases with increasing the temperature (see Fig. 2(d)). In addition, as expected, for films with the same width and length but thinner thickness (8~nm  film) the $h/e$ Aharonov-Bohm (AB) oscillation becomes more prominent (see Fig. 2(e)).  The observation of the  AAS and/or AB effect attests for the fact that the films under study have an insulating interior but a metallic surface. In particular, in order for the threaded flux to affect transport not only the top and the bottom  but also the side surfaces need to be metallic.  These metallic surface states are apparently very robust as they survived the ``brutal'' e-beam lithography.

Besides the AB and AAS oscillations, a careful examination of Fig. 2(c) reveals other finer structures (e.g., the features between peaks -5 and -6, and 2 and 3).   We attribute them to the universal conductance fluctuations (UCF) \cite{UCF}. Fig. 2(f) shows a couple of magneto-conductance curves (with the magnetic field applied perpendicular to the Bi film)  after background subtraction. For the red trace the magnetic field sweeps from -4 to 4 T, while for the blue trace the magnetic field is reversed (see the arrows). All the fine structures in Fig. 2(f) (with the fluctuation amplitude $\sim e^2/h$) are completely reproducible. They are the consequence of quantum interference of the electron matter waves in disordered media.

\begin{figure}
\includegraphics[width=8.5cm]{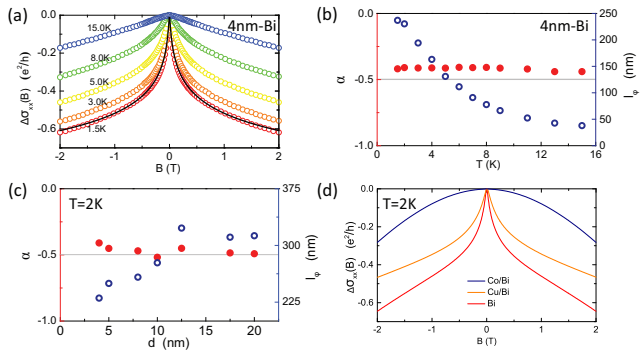}
\caption{\label{fig3}(a) Representative magneto-conductivity curves for 4~nm Bi film with perpendicular magnetic field at varies temperature, the black line shows the fitting with the HNL equation. (b) $\alpha$ and $l_{\varphi}$ extracted form the HNL equation fitting plotted as a function of temperature. (c) $\alpha$ and $l_{\varphi}$ at 2~K plotted as a function of thickness.(d) Impurity effect on the weak anti-localization.}
\end{figure}

Next we focus on the weak anti-localization (WAL) peak. In Fig. 3(a) we show the magnetoresistance curves for several different temperatures. These curves are measured on a 4nm thick Bi film patterned into a Hall bar. The WAL peak is clearly observed. After detailed quantitative analysis we conclude that these magnetoresistance data are well described by the two dimensional Hikami-Larkin-Nagaoka (HLN) theory for transport  in the presence of strong spin-orbit coupling
\begin{equation}
\Delta\sigma_{xx}(B)
=\alpha {e^{2}\over 2\pi^{2}\hbar}\Big[\Psi\Big({1\over 2}+{\hbar\over 4eBl_{\varphi}^{2}}\Big)-\ln\Big({\hbar\over 4eBl_{\varphi}^{2}}\Big)\Big],
\label{hln}\end{equation}
where $\Delta\sigma_{xx}(B)=\sigma_{xx}(B)-\sigma_{xx}(0)$,  $\Psi$ is the digamma function \cite{HNL}. Treating $\alpha$ (see later) and the phase coherence length $l_{\varphi}$ as fitting parameters we fit the data to \Eq{hln}. An example of the quality of the fit is shown by the bottom data set (the black curve is the fit) of Fig. 3(a). In  Fig. 3(b) we plot the fitted $\alpha$ value (red dots) as a function of temperature, the result suggests $\alpha$ is nearly temperature independent and is very close to -1/2. In contrast the fitted phase coherence length (blue dots) decreases monotonically with increasing temperature. In Fig. 3(c) we plot the fitted $\alpha$ as a function of film thickness for a fixed temperature. Again $\alpha$ is approximately thickness independent and stays very close to -1/2.
Theoretically $\alpha$=~-1/2 is expected for a connected spin-orbit interacting 2D system \cite{Alpha}. The fact that our fitted value is close to -1/2 again attests for the fact that the metallic surface states of our Bi film (top, bottom, and side surfaces together) forms a continuous 2D spin-orbit metal as expected for the surface states of topological insulator.

In Fig. 3(d) we show the effects of scalar and magnetic impurity on WAL. From the line shape of the curves it is clear that the WAL peak remains as a 0.5 ML of non-magnetic Cu are introduced to the top surface. In contrast when we add a 0.25 ML of magnetic Co the WAL peak is completely gone. This result supports the notion that the surface states of our Bi film are protected by the time reversal symmetry just like the surface states of topological insulators.

\begin{figure}
\includegraphics[width=4.5cm]{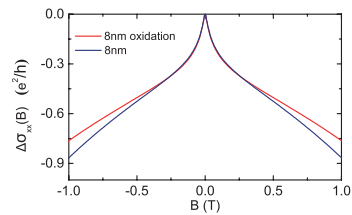}
\caption{\label{fig4}  Magneto-conductivity for 8 nm Bi film with and without oxidation.}
\label{ox}
\end{figure}

Before turning to theory, we checked that the metallic surface states of Bi(111) thin films are robust against surface oxidation. In growing  an 8~nm thick  Bi thin film we deliberately expose half of the sample to air (at 300~K) before capping. Interestingly such ``violent act'' only caused a merely 10 percent change in the total conductance. In Fig.~\ref{ox} we show the effect of oxidization on the magnetoresistance curve in weak field. Clearly the WAL peak is robust against oxidization. These results are consistent with the picture that surface oxidation simply displaces the metallic surface toward the film interior.

\begin{figure}
\includegraphics[scale=0.9]{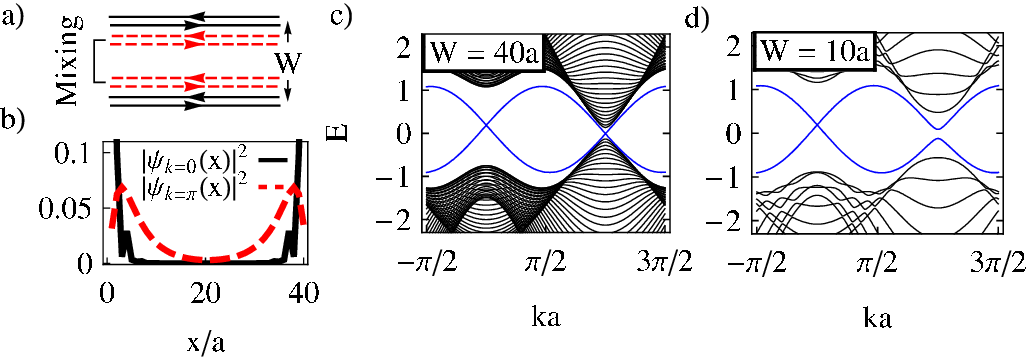}
\caption{(a) Schematic illustration of the physics which allows a trivial bulk system to become
topologically non-trivial when it is made into a thin film. (b) For the model
discussed in the text, wavefunctions of the edge modes at $k=0$ and $\pi$ for $W$=40$a$. (c) Spectrum
for $W$=40$a$. There are two pairs of counter-propagating Kramers pairs and hence the system is
topologically trivial (note every mode is doubly degenerate). (d) Spectrum of the same insulator
with $W$=10$a$. Only one pair of edge states remain gapless.}
\label{ribbon}
\end{figure}

All the experimental data presented up to this point supports the idea that when Bi (which is topologically trivial in the bulk) is made into thin films they become topological insulators. Here we demonstrate, as a matter of principle, that a topologically trivial system can become topologically non-trivial as it becomes sufficiently thin. For simplicity we focus on a two dimensional system (the generalization to 3D is straightforward). The basic physics is illustrated in Fig.~\ref{ribbon}(a) where each edge possesses two pairs of counter propagating helical modes (black and red) when $W$ is wide. With impurities backscattering between the red and black modes is allowed, hence
the system is topologically trivial. However if the red edge modes decay into the bulk with a much
larger length scale than the black ones there will be an intermediate range in $W$ where
backscattering between the red edge modes from opposite edges is sufficiently strong but that
between the black ones is still extremely weak. For that range of $W$ the red modes will be gapped
out and the system becomes indistinguishable from a strong topological insulator. The above physics
can be demonstrated by an explicit model
\be H = \left(\begin{smallmatrix}
           H_0(\v k) & 0 \\
           0 & H_0^*(-\v k)
          \end{smallmatrix}\right), {\rm where~} H_0(\v k) = \epsilon_{\v k} + \vec
\sigma \cdot \vec d(\v k).\ee  Here $\vec\sigma$ are the Pauli matrices, $\epsilon_{\v k} = A(\cos k_x +\cos k_y)$, $d_x + i d_y = (\sin k_x + i
\sin k_y)$ and $d_z = M - 4+2(1-\delta)\cos k_x +2(1+\delta)\cos k_y]$. For $A=0.3,M=3.3,
\delta=0.2$, this model describes a trivial insulator, featuring two gapped Dirac cones at $k=(0,\pi)$ and
$k=(\pi,0)$ with masses $m_{(0,\pi)} \gg m_{(\pi,0)}$. For large $W$ (e.g., $W$=40$a$),
this model has two pairs of counter-propagating edge modes centered at edge momenta $k=0$ and $\pi$,
see Fig.~\ref{ribbon}(c). Due to the
difference in the masses the decay length of these two edge modes satisfy $L_0\propto m_{(0,\pi)}^{-1}<< L_{\pi}\propto m_{(\pi,0)}^{-1}$ as Fig.~\ref{ribbon}(b) shows. Reducing $W$ makes states from opposite edges hybridize. However if $L_0 < W << L_\pi$, only the $k=\pi$ edge states hybridize significantly, causing them to open a gap (Fig.~\ref{ribbon}(d)). After that only one pair of edge states remain gapless and the system is indistinguishable from a strong topological insulator.

It is well known that bulk Bi is topologically trivial but Sb is topologically non-trivial. The difference between the bulk band structure of these two materials lies in the fact that gap inversion occurs at three time reversal invariant momenta, the L points, for Bi but not Sb. It is also known that the spin orbit induced gap at L points in Bi is very small ($\sim 15$ meV) \cite{GAPL} while the gap is large ($\sim 0.72-0.81$ eV) at $\Gamma$ point \cite{GAP1,GAP2}. They play the roles of the small and large Dirac masses respectively in the simple model above. To further check the existence of topological thin film state discussed above we have performed first-principle calculations for a superlattice made up of Bi slabs of varying thickness stacked in the (111) direction (we also can tune the coupling between adjacent slabs by adjusting the vacuum thickness between slabs). The result show that for both 12~BL and 15~BL Bi(111) slabs the artificial 3D system is topologically non-trivial because the gap at the L points becomes de-inverted. The detailed results will be presented elsewhere \cite{Wu}.

In summary we have carried out extensive quantum transport measurements and performed simple model as well as first-principle calculations for Bi thin films. All of our results suggest that when made thin enough Bi becomes topologically non-trivial. We explain this result in terms of a simple physical picture. We stress that what we discuss here is entirely different from the predicted two dimensional topological insulator when Bi film is only one bilayer thick. Our result should point out a new direction in the search of topologically interesting materials.\\

\begin{acknowledgments}
The authors thank S. Bl\"{u}gel,  G. Bihlmayer and I. Aguilera for their valuable discussions.
XFJ is supported by MOST (No. 2011CB921802) and NSFC (No. 11374057). DQ is supported by NSFC ( No. 11274228). JF is supported by NBRPC(2013CB921900).
DHL and FdJ are supported by DOE Office of Basic Energy Sciences, Division of Materials Science, Material theory Program, grant DE-AC02-05CH11231.
\end{acknowledgments}

\subsection{}
\subsubsection{}

\end{document}